\documentclass[conference]{IEEEtran}
\IEEEoverridecommandlockouts

\PassOptionsToPackage{hyphens}{url}\usepackage{hyperref}
\usepackage{hyperref}

\usepackage{cite}
\usepackage{amsmath,amssymb,amsfonts}
\usepackage{algorithmic}
\usepackage{graphicx}
\usepackage{textcomp}
\usepackage{xcolor}
\usepackage{float}
\usepackage{subcaption}
\floatstyle{ruled}
\newfloat{model}{thp}{lop}
\floatname{model}{Model}
\def\BibTeX{{\rm B\kern-.05em{\sc i\kern-.025em b}\kern-.08em
    T\kern-.1667em\lower.7ex\hbox{E}\kern-.125emX}}

\begin{document}

\title{
Evaluating Undergrounding Decisions for Wildfire Ignition Risk Mitigation across Multiple Hazards

}

\author{
\IEEEauthorblockN{Ryan Piansky and Daniel K. Molzahn}
\IEEEauthorblockA{\textit{School of Electrical and Computer Engineering} \\
\textit{Georgia Institute of Technology}\\
Atlanta, GA, USA \\
\{rpiansky3, molzahn\}@gatech.edu}
\and
\IEEEauthorblockN{Nicole D. Jackson}
\IEEEauthorblockA{\textit{Climate Security Center} \\
\textit{Sandia National Laboratories}\\
Albuquerque, NM, USA \\
njacks@sandia.gov}
\and
\IEEEauthorblockN{J. Kyle Skolfield}
\IEEEauthorblockA{\textit{Center for Computing Research} \\
\textit{Sandia National Laboratories}\\
Albuquerque, NM, USA \\
jkskolf@sandia.gov}
}

\maketitle

\begin{abstract}
With electric power infrastructure increasingly susceptible to impacts from climate-driven natural disasters, there is an increasing need for optimization algorithms that determine where to harden the power grid. Prior work has primarily developed optimal hardening approaches for specific acute disaster scenarios. Given the extensive costs of hardening the grid, it is important to understand how a particular set of resilience investments will perform under multiple types of natural hazards. Using a large-scale test case representing the Texas power system, this paper aims to understand how line undergrounding investment decisions made for wildfire ignition risk mitigation perform during a range of wildfire, hurricane, and wind events. Given the varying geographical spread and damage profile of these events, we show that investment decisions made to address one type of natural disaster do not necessarily improve broader resilience outcomes, supporting the need for co-optimization across a range of hazards. 
\end{abstract}


\section{Introduction}
The power grid faces increased frequency of weather-related power outages as climate change worsens~\cite{kenward2014blackout}. For instance, extreme weather related outages have had an average annual cost of \$127.4 billion from 2019 to 2024 in the United States~\cite{noaa2024billion}. Sustained power outages, largely attributable to extreme weather events~\cite{climate2024weather}, had an annual cost of over \$40 billion in 2015, a 25\% increase from 2002~\cite{lacommare2018improving}. 
Parallel to the frequency and severity of natural disasters, investments in climate resilience are growing. A 2024 Biden-Harris initiative announced \$2 billion for improved resilience to mitigate impacts from hurricanes, tornadoes, wildfires, and other extreme weather events~\cite{energy2024biden}. This investment is part of a larger \$100~billion in spending on power infrastructure, resilience, and clean energy~\cite{whitehouse2024fact}. The initiative in~\cite{energy2024biden} specifically includes funding for projects in Texas, which has seen power outages from hurricanes~\cite{berg2009tropical}, wildfires~\cite{powerfiretexas}, ice storms~\cite{levin2022extreme}, and strong winds~\cite{phillip2024some}. One study found the system average interruption duration index (SAIDI) in the West South Central region (including Texas) increased 6\% annually from 2000 to 2015, a finding which is ``correlated with high wind speeds''~\cite{larsen2020severe}. As such, Texas provides an interesting case study due to its large geographic area and exposure to a wide variety of extreme weather events that are sometimes localized within a given region but can extend across large portions of the state. 

As investment for climate resilience expands, researchers are investigating methods to optimally design and operate transmission-scale infrastructure~\cite{garland2024climate,bhusal2020power}. 
Prior research, e.g.,~\cite{bynum2021proactive}, considers long-term investment planning with Texas case studies; however, this research largely considers random outages without using actual geographic weather data.
Research has also been performed investigating cascading failures from line outages under extreme hurricane events in Texas, suggesting power lines be hardened or undergrounded to remove the likelihood of failure~\cite{sturmer2024increasing}. Many research efforts have evaluated optimal infrastructure decisions for specific climate-related resilience including winterization of generators in Texas with consideration for equity~\cite{bilir2024enhancing} or winterization through mobile battery systems on a reduced Texas system~\cite{austgen2024three}. Similarly, reference~\cite{pollack2024equitably} considers equitable line undergrounding for wildfire ignition risk mitigation. Research in \cite{piansky2024long} investigates large-scale battery installation to support both wildfire-related power outages and nominal operations. 

Outside of Texas, researchers have considered power system planning in climate-related contexts. Reference~\cite{skolfield2021transmission} investigates expansion planning for transmission systems under increasing temperatures in Arizona. Reference~\cite{nasri2022multi} develops a multi-stage stochastic distribution expansion model for resilience under hurricane-related vulnerabilities. Likewise, reference~\cite{newlun2019co} formulates a capacity expansion planning model for hurricane resilience in Puerto Rico. Additionally, reference~\cite{moreira2021climate} proposes a three-stage robust transmission expansion problem under climate uncertainty in Chile. 

Wildfires pose a unique climate event since power infrastructure can not only be damaged by active wildfires but also can ignite wildfires under conditions of high temperatures, high winds, and low humidity~\cite{CPUCignitions}. To combat this, utilities conduct Public Safety Power Shutoff (PSPS) events to mitigate wildfire ignition risk, but these events can lead to load shedding for consumers~\cite{sotolongo2020california}. Distributed energy resources are one solution evaluated in~\cite{bayani2023resilient}. Similarly, networked microgrids are explored in~\cite{Taylor2023Managing}. Line undergrounding for PSPS events has also been extensively explored~\cite{kody2022optimizing,taylor2023robust,taylor2022framework}. Undergrounded lines effectively remove the risk of igniting a wildfire while allowing these lines to continue transmitting power. Undergrounded lines are also less susceptible to other extreme events, such as winds, hurricanes, or active fires.

This paper evaluates the performance of climate resilience investment decisions developed for a specific type of extreme weather event under other types of natural disasters. In particular, we consider real, extreme weather-driven power outage data associated with a large-scale and realistic synthetic transmission grid geolocated in Texas~\cite{xu2017creation}. We evaluate how line undergrounding decisions made to mitigate wildfire ignition risk impact power outages under three types of extreme weather events: hurricanes, wind, and active wildfires from actual 2021 data. To the best of our knowledge, this is the first power systems paper to analyze the impacts of infrastructure investments made to mitigate one hazard in the context of a range of other hazards for a large-scale realistic test case.  

\section{Data}

In this paper, we evaluate real-world data from 2021 geo-located in Texas for wildfires, hurricanes and wind based events. We augment and evaluate this data with an undergrounding plan devised to reduce load shed during PSPS events for wildfire ignition risk mitigation.  

\subsection{Outage Data}\label{sec:outages}

County-level grid outage information (i.e., number of customers affected) was obtained from the Environment for Analysis of Geo-Located Energy Information (EAGLE-I) dataset~\cite{brelsford2024dataset} for the year 2021. The outage information was converted to an hourly resolution and further refined to identify discrete, multi-hour outage events. The outages were paired with weather event data from the National Oceanic and Atmospheric Administration (NOAA) National Centers for Environmental Information (NCEI) storm event database~\cite{noaa2024sed}. For this work, we consider three weather-based hazards: wind (i.e., high winds, strong winds, and thunderstorm winds); hurricanes; and wildfires. The combined data provides information on the average and maximum number of customers affected during an outage, total duration, and binary indicators for the presence of a storm event.

\subsection{Undergrounding Plan}\label{sec:ug_plan}
\begin{figure}[t]
    \centering
    \includegraphics[width=\linewidth,trim={0 6em 0 2em},clip]{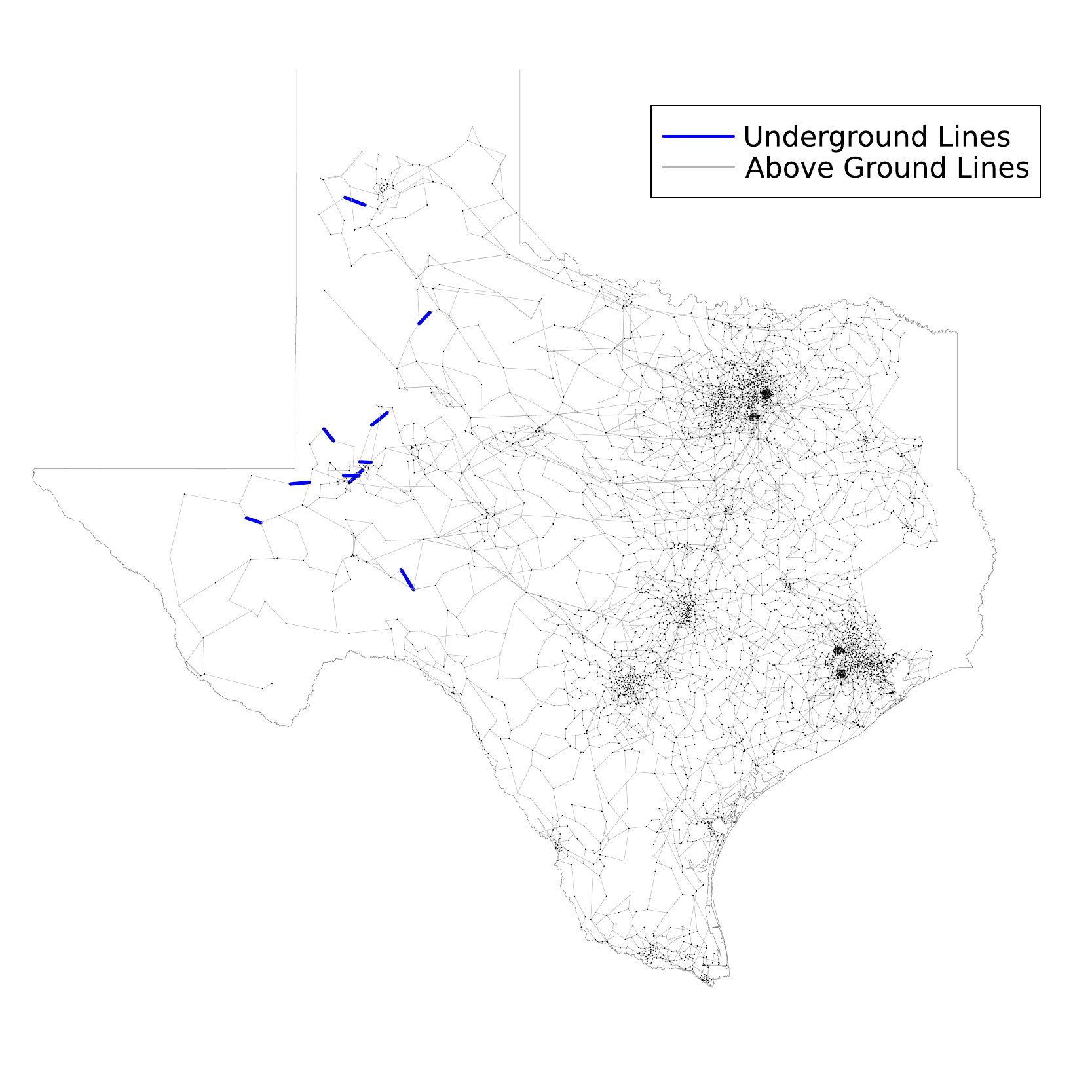}
	\caption{A map of the line undergrounding plan considered in this paper based on~\cite{pollack2024equitably}. Grey lines show above ground transmission lines (at risk for outages). Bold blue lines represent planned transmission lines for undergrounding.}
	\label{fig:ug_lines}
    \vspace{-1em}
\end{figure}

We consider wildfire ignition risk mitigation via the undergrounding investment plan from~\cite{pollack2024equitably}. In this paper, a budget of \$1 billion USD is used to underground transmission lines in Texas to minimize the amount of load shedding associated with PSPS events. The authors in~\cite{pollack2024equitably} found the undergrounding investment plan to reduce the load shed from PSPS events by over 70\%. These events proactively de-energize transmission lines that are at high risk for igniting wildfires through faults caused by certain extreme weather conditions~\cite{huang2023review}. While expensive, undergrounding transmission lines allows power to continue flowing while removing the risk of that line igniting a fire. The authors in~\cite{kody2022optimizing} find that line undergrounding is a very effective wildfire ignition risk mitigation option when compared to other line hardening methods (vegetation management or covered conductors) or infrastructure such as solar PV installation or grid-scale batteries. The planned lines for undergrounding are shown as the bold blue lines in Figure~\ref{fig:ug_lines}, largely in northwestern Texas. For more information on how these lines are optimally selected for undergrounding, see~\cite{pollack2024equitably}. 

\subsection{Test Network}

This research uses the Texas7k network, a synthetic transmission network developed by the Texas A\&M PERFORM group~\cite{xu2017creation}. With 6717 buses, 9140 transmission lines, and 731 generators, this network provides a realistic transmission system covering the area in ERCOT~\cite{xu2018synthetic}.

\section{Methodology}

We generate 100 outage scenarios for each day of the year for the three discussed hazards: wildfires, hurricanes, and wind events. We evaluate the load shed in a pre-resilience case, i.e., when all outaged lines are unable to transmit power. We also evaluate the post-resilience case where the undergrounded lines that would otherwise have been outaged remain energized and can transmit power.

\subsection{Scenario Development}\label{sec:scenario}

Using the combined datasets, we generated 100 different outage scenarios per event type per day. The dataset provides the fraction of customers experiencing load shed in each county as well as a binary hazard indicator. We take the following approach to determine the sets of line outages for each event in each scenario:
\begin{enumerate}
    \item $\rho_{d,c} = \sum_{t \in \mathcal{T}} \rho_{d,c,t}$ --- the outage probability proxy for a county $c \in \mathcal{C}$ on day $d$ where $\rho_{d,c,t}$ is the proportion of customers experiencing an outage in county $c$ at time~$t$ on day $d$.
    \item $o_{l,d,k}$ --- a random outage number for each line $l \in \mathcal{L}$ in scenario $k$ chosen from a beta prime distribution with a mean value of 0.01~\cite{mahmood3analysis,mahmood3utilizing}.
    \item For each line $l$, we compare $o_{l,d,k}$ to $\rho_{d,c}$, $\forall c \in \mathcal{C}_l$, the set of counties intersected by line $l$. If $o_{l,d,k} < \rho_{d,c}$, line~$l$ is outaged (included in the set $\mathcal{L}^\text{out}_{d,k}$) for the corresponding day. 
\end{enumerate}

\begin{table}[t]
\begin{center}
\begin{tabular}{|c|c|c|c|c|}
\hline
          & \textbf{Avg Outages} & \textbf{Max Outages} & \textbf{Days} & \textbf{Scenarios} \\ \hline
\textbf{Wildfire}  & 13.2 & 20 & 2 & 109 \\ \hline
\textbf{Hurricane} & 900.3 & 1230 & 2 & 200 \\ \hline
\textbf{Wind}      & 24.7 & 303 & 86 & 6239 \\      
\hline
\end{tabular}
\end{center}
\caption{Statistics for line outages co-occurring with the three examined types of weather events. The average outages column indicates the average number of lines outaged across all scenarios on all days (excluding scenarios with no outages) for each event type. The max outages column indicates the maximum number of lines outaged across all scenarios on all days. The days column indicates the total number of days where any outages occur. The scenarios column indicates the number of scenarios across all days where any outages occur.}
\label{tab:lines_outaged}
\vspace{-1em}
\end{table}

Table~\ref{tab:lines_outaged} shows summary statistics that capture the average and maximum number of line outages are across all scenarios and days. Only scenarios with at least one line outage are counted when finding the average. Wildfires see relatively few lines outaged, with roughly 13 lines outaged on average across the scenarios. Hurricanes see an order of magnitude more outages with over 900 lines being outaged on average across the scenarios. Wind scenarios occur much more often than both hurricanes and wildfires and much more extensively throughout the year, causing almost twice the number of line outages on average.

We see extensive outages co-occurring with hurricanes, specifically on September 13/14\textsuperscript{th}. On these days, a large portion of the synthetic transmission network is outaged. Similarly, wildfire outages also only occur on two days (April 10\textsuperscript{th} and December 10\textsuperscript{th}). However, there are only reported power outages in a few counties, leading to very few lines being impacted in our scenarios. Wind-related events occur much more sporadically throughout the year, impacting dozens to hundreds of lines, during related power outages. 

\begin{figure}[htbp!]
    \centering
    \vspace{-2em}
	\includegraphics[width=\linewidth,trim={0 7em 0 3em},clip]{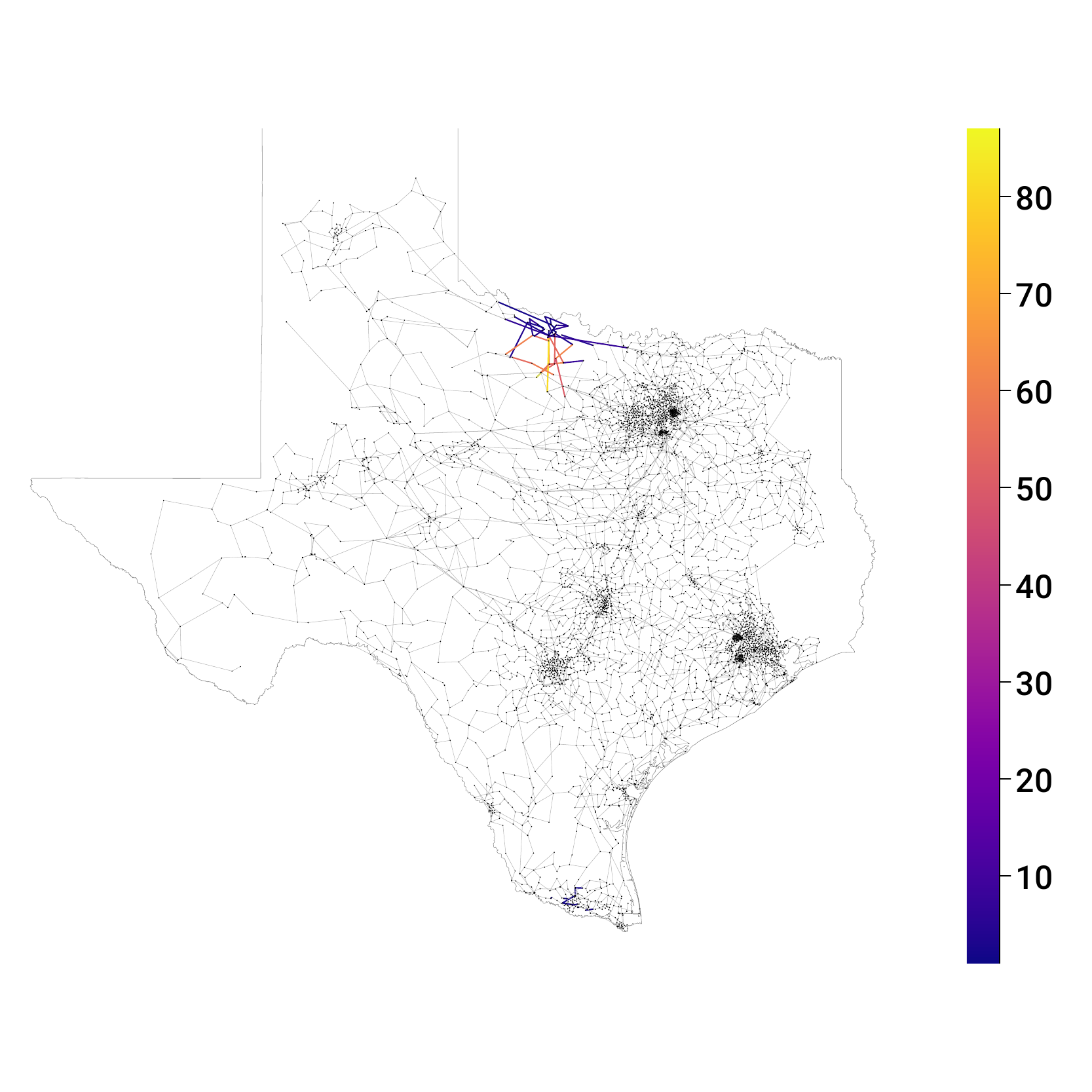}
	\caption{A map showing line outages which co-occurred with wildfires. The most-outaged line (in yellow) is shutoff 87 times across all days and all scenarios while the least-outaged line (in purple) was only outaged 1 time.}
	\label{fig:wf_outs}  
\end{figure}

\begin{figure}[h]
    \centering
    \vspace{-2em}
	\includegraphics[width=\linewidth,trim={0 7em 0 1em},clip]{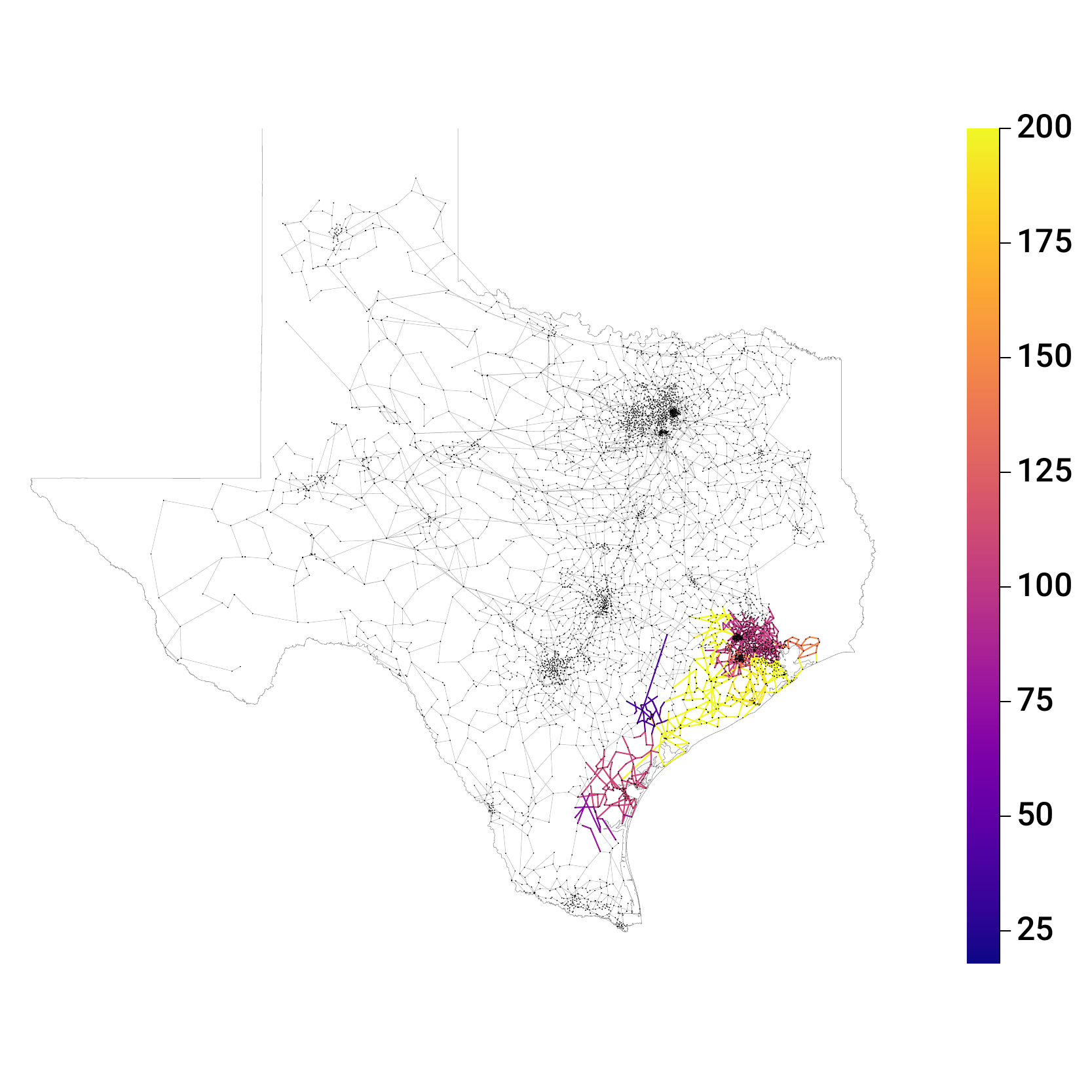}
	\caption{A map showing line outages which co-occurred with hurricanes. The most-outaged line (in yellow) is shutoff 200 times across all days and all scenarios while the least-outaged line (in purple) was only outaged 18 times.}
	\label{fig:hurricane_outs}  
    \vspace{-1em}
\end{figure}

\subsection{Power System Model}\label{sec:model}

To evaluate the load shedding in each scenario on each day, we run a B$\theta$ DC-OPF model with an objective to minimize total network load shed. We define $\mathcal{L}^{\text{out}}_{d,k}$ as the set of outaged lines on a given day and scenario for the outage types discussed. We model hourly operations on each day, with hourly demand information adapted from~\cite{rts96}. The mathematical model is shown in Model~\ref{model:opf}, $\forall d \in \mathcal{D}, \forall k \in \mathcal{K}$, where $\mathcal{D}$ is the set of days and $\mathcal{K}$ is the set of scenarios.

\begin{model}[h]
\caption{B$\theta$ DC-OPF}
\label{model:opf}
\begin{subequations}
\vspace{-0.2cm}
\begin{align}
& \mbox{\textbf{min}} \  \sum_{t \in  \mathcal{T}}\sum_{n \in \mathcal{N}} p_{ls,t}^n\label{eq:obj} 
\\
& \mbox{\textbf{s.t.}}  \quad \forall t \in \mathcal{T}, \nonumber \\
& \underline{p}_{g}^i \leqslant p_{g,t}^i \leqslant \overline{p}_{g}^i 
&& \hspace{-4.1em} \forall i \in \mathcal{G} \label{subeq: gen limits} 
\\
& 0 \leqslant p_{ls,t}^n \leqslant p_{l,t,d}^n 
&& \hspace{-4.1em} \forall n \in \mathcal{N} \label{subeq: loadshed limits} 
\\
& f^\ell_{t} = 0
&& \hspace{-4.1em} \forall \ell \in \mathcal{L}^{\text{out}}_{d,k} \label{subeq: power off} 
\\
& -\overline{f}^\ell \leqslant f^\ell_{t} \leqslant \overline{f}^\ell 
&& \hspace{-4.1em} \forall \ell \in \mathcal{L} \setminus \mathcal{L}^{\text{out}}_{d,k} \label{subeq: power flow limits} 
\\
& \underline{\delta}^\ell  \leqslant \theta^{n^{\ell, \text{fr}}}_{t} \!\!\!\! - \theta^{n^{\ell, \text{to}}}_{t} \leqslant \overline{\delta}^\ell 
&& \hspace{-4.1em} \forall \ell \in \mathcal{L}\setminus\mathcal{L}^{\text{out}}_{d,k} \label{subeq: voltage angle} 
\\
& f_t^\ell = -b^\ell(\theta^{n^{\ell, \text{fr}}}_{t} \!\!\!\! - \theta^{n^{\ell, \text{to}}}_{t}) 
&& \hspace{-4.1em}  \forall \ell \in \mathcal{L}\setminus\mathcal{L}^{\text{out}}_{d,k} \label{subeq: power flow}
\\
& \! \sum_{\ell \in \mathcal{L}^{n, \text{fr}}} \!\!\!\! f^\ell_{t} \! - \!\!\!\! \sum_{\ell \in \mathcal{L}^{n, \text{to}}} \!\!\!\! f^\ell_{t} \! = \!\!\!  \sum_{i \in \mathcal{G}^n} \!\! p_{g,t}^i \! - \! p_{l,t,d}^n \! + \! p_{ls,t}^n 
&& \hspace{-1.8em} 
\forall n \in \mathcal{N} \label{subeq: power balance}
\end{align}
\end{subequations}
\end{model}

Objective~\eqref{eq:obj} minimizes the sum of load shed across all buses and time periods. Equation~\eqref{subeq: gen limits} constrains the power generation ($p_{g,t}^i$) to be between the lower and upper limits for each generator $i \in \mathcal{G}$, the set of all generators. Equation~\eqref{subeq: loadshed limits} constrains the load shed ($p_{ls,t}^n$) to be nonzero and less than the load demanded ($p_{d,t}^n$) at the same time and for each bus $n \in \mathcal{N}$, the set of buses. Equation~\eqref{subeq: power off} ensures the power flow ($f_{t}^{\ell}$) is zero for any line $\ell \in \mathcal{L}^\text{out}_{d,k}$, the set of outaged lines on day $d$ in scenario $k$. For any lines that remain operational, equation~\eqref{subeq: power flow limits} constraints the power flow to be between the lower and upper limits. Equation~\eqref{subeq: voltage angle} constrains the voltage angle difference across a line to be within the lower and upper bounds, where $\theta^{n^{l,fr}}_{t}$ and $\theta^{n^{l,to}}_{t}$ are the voltage angles associated with the ``from'' and ``to'' bus of line $l$, respectively. Equation~\eqref{subeq: power flow} models the B$\theta$ power flow approximation for each line. Finally, equation~\eqref{subeq: power balance} models the power balance constraints. For the post-resilience results discussed in Section~\ref{sec:post_out}, the set of outaged lines is defined as $\mathcal{L}^{out}_{d,k} \setminus \mathcal{L}^{ug}$ where $\mathcal{L}^{ug}$ is the set of undergrounded lines shown in Figure~\ref{fig:ug_lines}.

The problem described in Model~\ref{model:opf} optimizes over a set of decision variables. An operating setpoint is defined over the decision variables tied to generation, voltage angles, and line flows at each time step; $p^{i}_{g,t}$, $\theta^{n}_{t}$, and $f^{\ell}_{t}$ respectively. In addition, the summation of the decision variable for load shed at each bus and time step, $p^{n}_{ls,t}$, is minimized in the objective.

\section{Event Overlap}\label{sec:overlap}
For the line undergrounding plan outlined in Section~\ref{sec:ug_plan}, we find there are no lines that are undergrounded for wildfire ignition risk mitigation that are outaged during the hurricane events or the active wildfire events from this study. For the wildfire events, the ignition risk data used for the undergrounding decisions comes from the United States Geological Survey's Wildland Fire Potential Index~\cite{USGS2024Wildland} which indicates where a significant fire is likely to \emph{start}. In contrast, the combined EAGLE-I and NOAA NCEI storm events dataset used for these outage scenarios show where \emph{active} wildfires are co-occurring with impacted power system operations. For the evaluated time period, outages from active wildfires are observed in the central-north part of the state (Figure~\ref{fig:wf_outs}), which does not overlap with the lines chosen for undergrounding. Similarly, for hurricanes, the lack of overlap can be explained by the geographic disparity between where lines are undergrounded due to high wildfire ignition risk (northwest Texas in Figure~\ref{fig:ug_lines}) and where the hurricane event captured by this research impact power infrastructure (southeast Texas in Figure~\ref{fig:hurricane_outs}).

\begin{figure}[t]
    \centering
	\includegraphics[width=\linewidth,trim={0 7em 0 10em},clip]{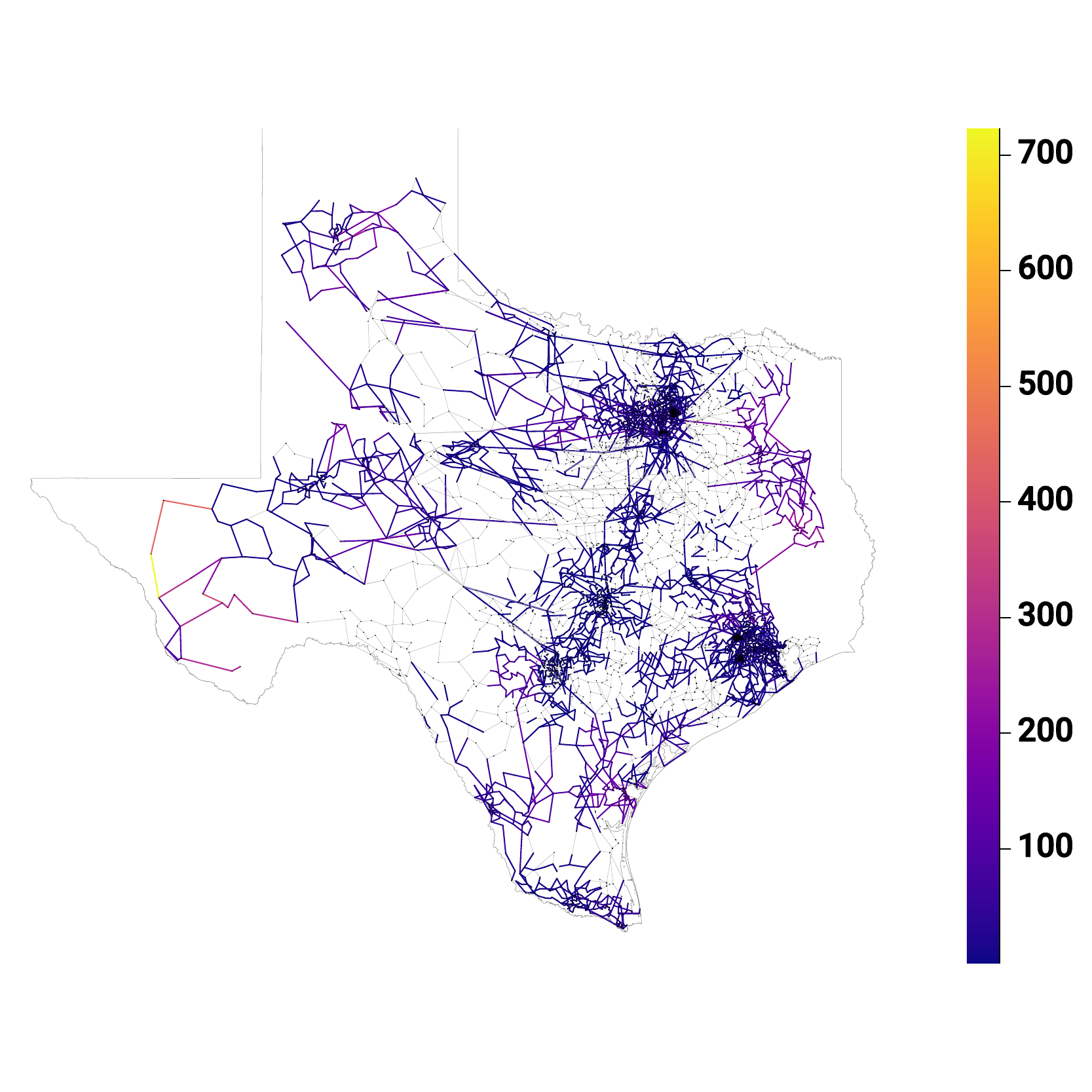}
	\caption{A map showing line outages which co-occurred with wind hazards. The most-outaged line (in yellow) is shutoff 723 times across all days and all scenarios while the least-outaged line (in purple) was only outaged 1 time.}
	\label{fig:wind_outs}  
    \vspace{-1em}
\end{figure}

Outage events co-occurring with wind events examined in this paper span a wide geographic area (see Figure~\ref{fig:wind_outs}). Some of the lines outaged from wind events overlap with those in the undergrounding plan. Of the 6,239 total scenarios with wind-related outages across the year-long time period examined, 252 scenarios involve lines that are undergrounded, or roughly 4\% of the wind-related outage scenarios. There are 12 total outage scenarios that are fully prevented, since all involved lines are undergrounded, representing less than 0.2\% of wind-related outage scenarios. Load shedding outcomes from pre- and post-undergrounding are discussed in Section~\ref{sec:results}.

\section{Results}\label{sec:results}

In this section, we show load shed outcomes for winds, hurricanes, and wildfire related outages without line undergrounding investments. As discussed in Section~\ref{sec:overlap}, there are no impacted lines in the hurricane or wildfire scenarios that are undergrounded in the wildfire ignition risk mitigation undergrounding plan. Thus, in Section~\ref{sec:post_out}, only results from wind events are shown to highlight the reduction in load shed when lines are undergrounded. Note that the figures in this section have different scales on the vertical-axis to better show detail for specific weather-based hazards. Load shed values in this sections are given in megawatt-hours (MWh).

\subsection{Pre-Resilience Outages}\label{sec:pre_out}

\begin{figure}[h]
    \centering
	\includegraphics[width=\linewidth]{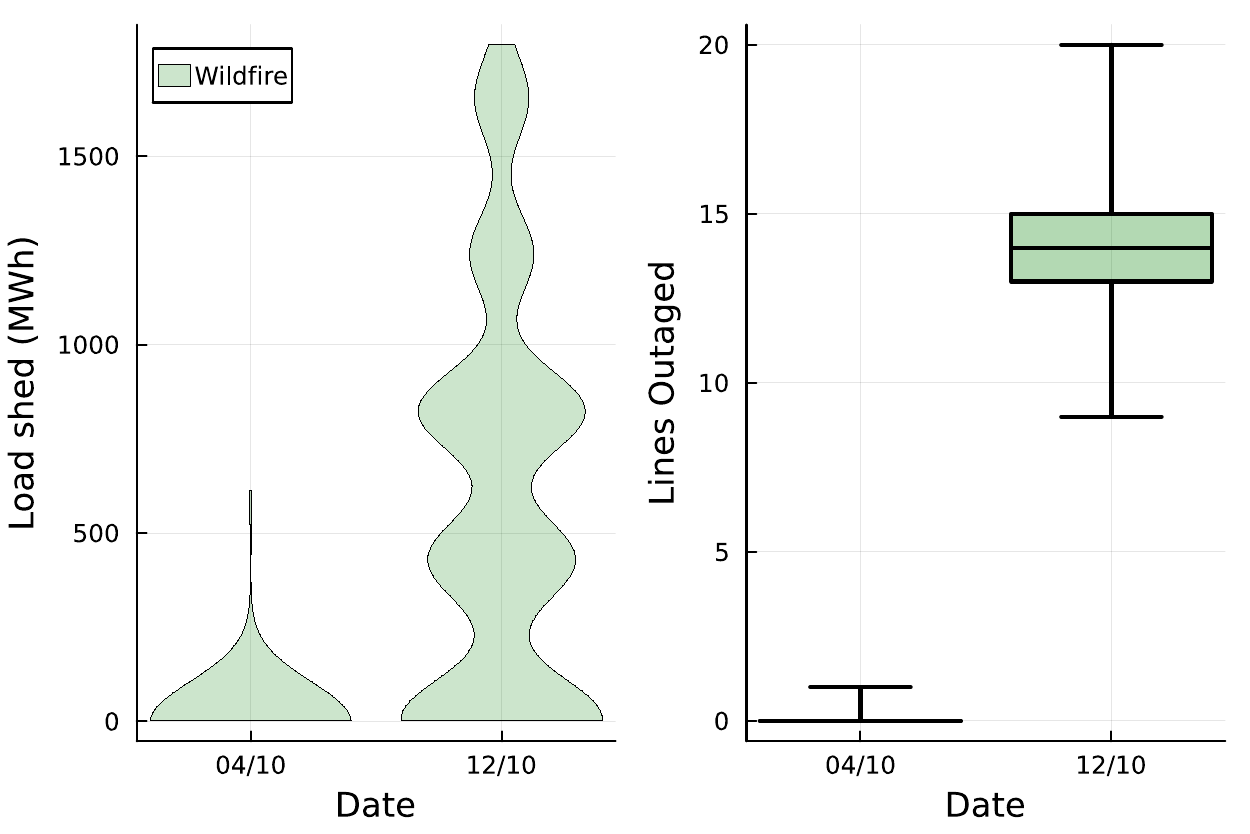}
	\caption{Violin plots show the daily load shed across scenarios on the two impacted days under the wildfire events. Box plots indicate the number of outaged lines co-occurring across the scenarios on a given day.}
	\label{fig:wf_ls}  
    \vspace{-1em}
\end{figure}

\begin{figure}[h]
    \centering
	\includegraphics[width=\linewidth]{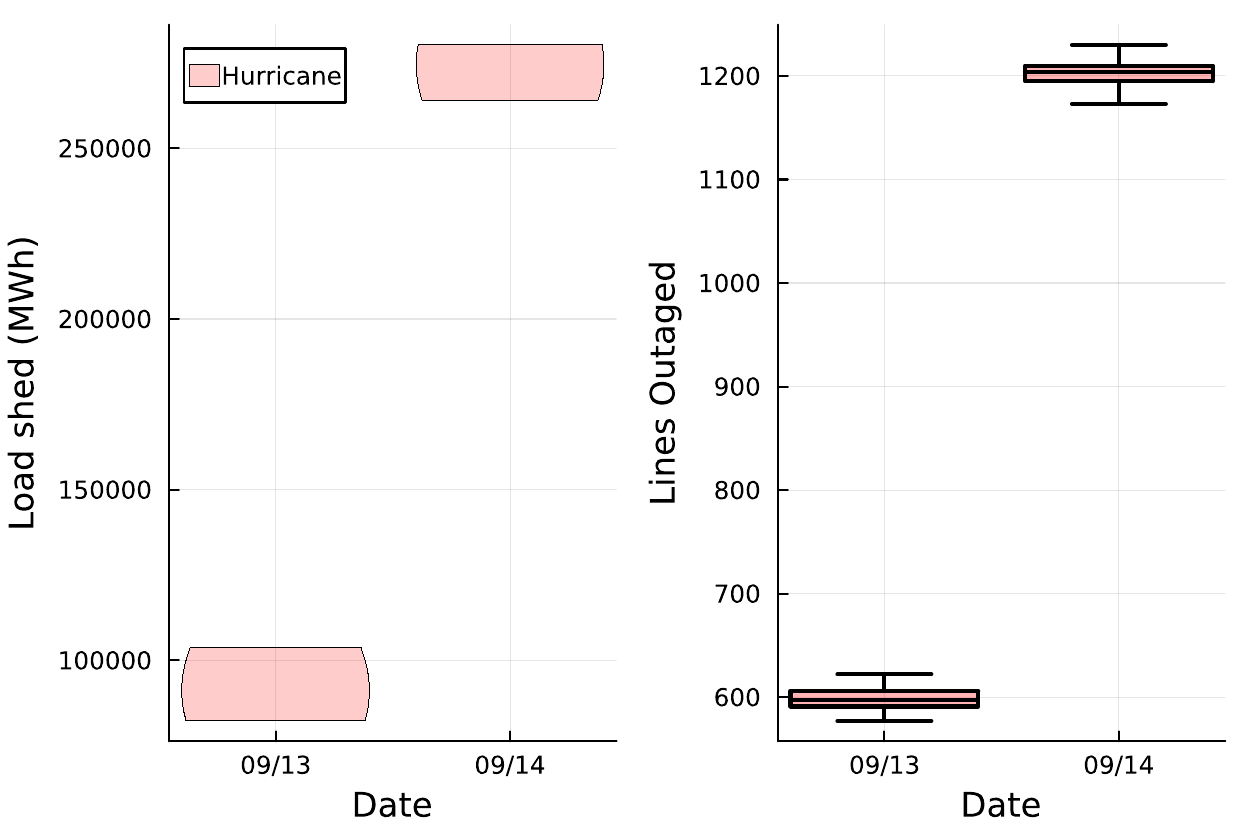}
	\caption{Violin plots show the daily load shed across scenarios on the two impacted days under the hurricane events. Box plots indicate the number of outaged lines co-occurring across the scenarios on a given day.}
	\label{fig:hurricane_ls}
    \vspace{-1em}
\end{figure}

\begin{figure}[h]
    \centering
	\includegraphics[width=\linewidth]{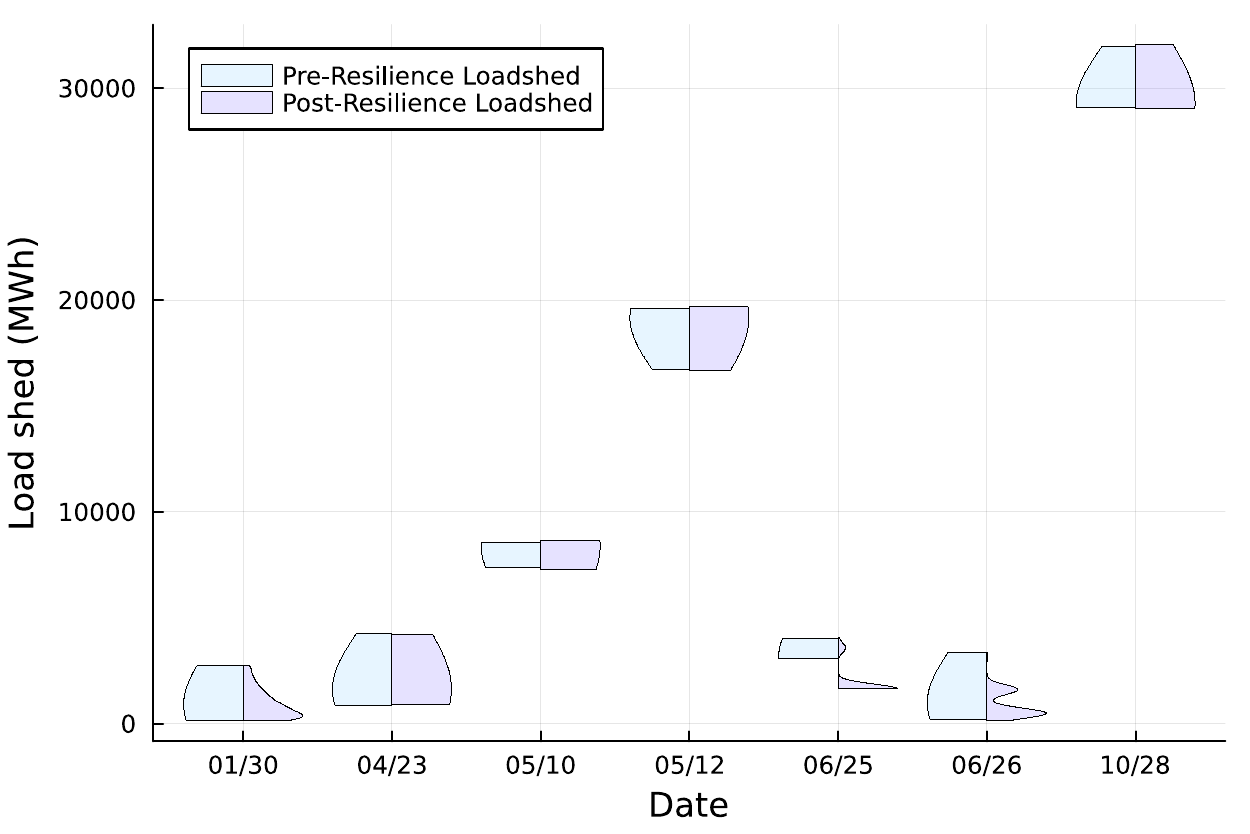}
	\caption{Violin plot showing the total amount of daily load shed across wind events on a subset of days. The left side shows pre-contingency load shed and the right side shows post-contingency load shed.}
	\label{fig:wind_ls}
    \vspace{-1em}
\end{figure}

As discussed in Section~\ref{sec:outages}, active wildfires caused very few impacts to the power grid during the evaluated time period. We see this again in Figure~\ref{fig:wf_ls} which shows around 1,000 MWh of load shed or less on December 10\textsuperscript{th} and even less on April 10\textsuperscript{th}. This constitutes less than 0.1\% of the total demand across the network on December 10\textsuperscript{th}. Likewise, we only see load shed with hurricanes on September 13/14\textsuperscript{th}, aligning with the limited line outages discussed in Section~\ref{sec:outages}. In Figure~\ref{fig:hurricane_ls}, we see that the hurricane-related outage events cause significant load shed, with the average load shed across the scenarios approaching 300,000 MWh on September 14\textsuperscript{th}. This constitutes nearly 25\% of the total demand on that day. Although the region impacted by the hurricane is a relatively small portion of Texas, as seen in Figure~\ref{fig:hurricane_outs}, the large number of line failures leads to widespread impacts across the network. On September 14\textsuperscript{th}, there are roughly 1200 lines outaged, comprising over 10\% of the synthetic transmission network, as shown by the box plots in Figure~\ref{fig:hurricane_ls}. 
As indicated in Figure~\ref{fig:wind_ls}, load shed throughout the considered time period varies from nearly zero to the order of 30,000 MWh of load shed on days with wind-related outages. Figure~\ref{fig:wind_ls} shows a subset of days with outages co-occurring with wind hazards. The maximum load shed that occurs is on October 28\textsuperscript{th}, representing less than 3\% of the demand on that day. 

\subsection{Post-Resilience Outages}\label{sec:post_out}

Wind-related line outages were the only type that intersected with the set of underground lines for wildfire ignition risk mitigation as discussed in Section~\ref{sec:overlap}. In Figure~\ref{fig:wind_ls}, we show the post-contingency load shed on the right side of the violin plots. Only three days saw load shed reduction when considering the lines undergrounded to support PSPS events: January 30\textsuperscript{th}, June 25\textsuperscript{th}, and June 26\textsuperscript{th}. The average decrease in load shed per scenario after considering undergrounded lines is 25MWh, 1250MWh, and 135MWh across the three days respectively. 


\section{Conclusion}
This paper compared line outage profiles of three types of extreme weather events based on real-world data correlated with a realistic synthetic transmission grid geolocated in Texas. Using an optimal power flow formulation, we computed the load shed during every hour of each day of 2021 with 100 scenarios per day. This provides a baseline to compare load shed results after we consider certain lines to be undergrounded and protected from the considered weather events. 

Line undergrounding decisions were based on their effectiveness in reducing load shed from PSPS events that mitigate the risk of wildfire ignitions from power infrastructure. While these undergrounding decisions improve resilience outcomes in this specific context~\cite{pollack2024equitably}, we find that they yield very small improvements for the other extreme events evaluated in this work, namely, active wildfires, hurricanes, and wind. The first two of these event types had no overlap between their associated line outages and the lines that were undergrounded for wildfire ignition risk mitigation. The third, wind events, had some overlap but only saw meaningful reduction in load shed on three days out of the year.

Thus, while the line undergrounding decisions are effective at reducing load shed for PSPS events that mitigate wildfire ignition risk, this paper shows that resilience investments planned and optimized to aid in a specific climate events can have negligible impacts on reducing climate impacts more broadly. This motivates further research on \emph{co-optimizing} infrastructure investments across a number of different geographically varying extreme weather events.

\section*{Acknowledgment}
This work was supported by the Laboratory Directed Research and Development program at Sandia National Laboratories, a multimission laboratory managed and operated by National Technology and Engineering Solutions of Sandia LLC, a wholly owned subsidiary of Honeywell International Inc. for the U.S. Department of Energy’s National Nuclear Security Administration under contract DE-NA0003525. This paper describes objective technical results and analysis. Any subjective views or opinions that might be expressed in the paper do not necessarily represent the views of the U.S. Department of Energy or the United States Government.

\bibliographystyle{IEEEtran}
\bibliography{IEEEabrv,refs.bib}

\begin{thebibliography}{10}
\providecommand{\url}[1]{#1}
\csname url@samestyle\endcsname
\providecommand{\newblock}{\relax}
\providecommand{\bibinfo}[2]{#2}
\providecommand{\BIBentrySTDinterwordspacing}{\spaceskip=0pt\relax}
\providecommand{\BIBentryALTinterwordstretchfactor}{4}
\providecommand{\BIBentryALTinterwordspacing}{\spaceskip=\fontdimen2\font plus
\BIBentryALTinterwordstretchfactor\fontdimen3\font minus \fontdimen4\font\relax}
\providecommand{\BIBforeignlanguage}[2]{{%
\expandafter\ifx\csname l@#1\endcsname\relax
\typeout{** WARNING: IEEEtran.bst: No hyphenation pattern has been}%
\typeout{** loaded for the language `#1'. Using the pattern for}%
\typeout{** the default language instead.}%
\else
\language=\csname l@#1\endcsname
\fi
#2}}
\providecommand{\BIBdecl}{\relax}
\BIBdecl

\bibitem{kenward2014blackout}
A.~Kenward, U.~Raja \emph{et~al.}, ``Blackout: Extreme weather, climate change and power outages,'' \emph{Climate Central}, vol.~10, 2014.

\bibitem{noaa2024billion}
\BIBentryALTinterwordspacing
{NOAA National Centers for Environmental Information (NCEI)}, ``{U.S. Billion-Dollar Weather and Climate Disasters},'' 2024. [Online]. Available: \url{{https://www.ncei.noaa.gov/access/billions/}}
\BIBentrySTDinterwordspacing

\bibitem{climate2024weather}
\BIBentryALTinterwordspacing
{Climate Central}, ``{Weather-related Power Outages Rising},'' 2024. [Online]. Available: \url{{https://www.climatecentral.org/climate-matters/weather-related-power-outages-rising}}
\BIBentrySTDinterwordspacing

\bibitem{lacommare2018improving}
K.~H. LaCommare, J.~H. Eto, L.~N. Dunn, and M.~D. Sohn, ``Improving the estimated cost of sustained power interruptions to electricity customers,'' \emph{Energy}, vol. 153, pp. 1038--1047, 2018.

\bibitem{energy2024biden}
{US Department of Energy}, ``{B}iden-{H}arris {A}dministration announces additional \$2 billion to protect the grid against growing threats of extreme weather, expand transmission,'' 2024.

\bibitem{whitehouse2024fact}
{The White House}, ``Fact sheet: {B}iden-{H}arris {A}dministration transforms nation’s infrastructure, celebrates historic progress in rebuilding {A}merica for the three-year anniversary of the {B}ipartisan {I}nfrastructure {L}aw,'' 2024.

\bibitem{berg2009tropical}
R.~Berg, \emph{Tropical cyclone report: Hurricane {I}ke (al092008), 1-14 {S}eptember 2008}.\hskip 1em plus 0.5em minus 0.4em\relax National Hurricane Center, 2009.

\bibitem{powerfiretexas}
{Texas A\&M}, ``How do power lines cause wildfires?'' \emph{Texas Wildfire Mitigation Project}, July 2014, \url{https://wildfiremitigation.tees.tamus.edu/faqs/how-power-lines-cause-wildfires}.

\bibitem{levin2022extreme}
T.~Levin, A.~Botterud, W.~N. Mann, J.~Kwon, and Z.~Zhou, ``Extreme weather and electricity markets: Key lessons from the {F}ebruary 2021 {T}exas crisis,'' \emph{Joule}, vol.~6, no.~1, 2022.

\bibitem{phillip2024some}
\BIBentryALTinterwordspacing
D.~Phillip, L.~Baumann, and C.~Weber, ``{Some Houston-area power outages could last weeks after deadly storms cause widespread damage},'' \emph{{The Texas Tribune}}, 2024. [Online]. Available: \url{https://www.texastribune.org/2024/05/16/texas-storm-deaths-houston-power-outage-damage/}
\BIBentrySTDinterwordspacing

\bibitem{larsen2020severe}
P.~H. Larsen, M.~Lawson, K.~H. LaCommare, and J.~H. Eto, ``Severe weather, utility spending, and the long-term reliability of the {US} power system,'' \emph{Energy}, vol. 198, 2020.

\bibitem{garland2024climate}
J.~Garland, K.~Baker, and B.~Livneh, ``The climate-energy nexus: A critical review of power grid components, extreme weather, and adaptation measures,'' \emph{Environmental Research: Infrastructure and Sustainability}, vol.~4, no.~3, September 2024.

\bibitem{bhusal2020power}
N.~Bhusal, M.~Abdelmalak, M.~Kamruzzaman, and M.~Benidris, ``Power system resilience: Current practices, challenges, and future directions,'' \emph{IEEE Access}, vol.~8, pp. 18\,064--18\,086, 2020.

\bibitem{bynum2021proactive}
M.~Bynum, A.~Staid, B.~Arguello, A.~Castillo, B.~Knueven, C.~D. Laird, and J.-P. Watson, ``Proactive operations and investment planning via stochastic optimization to enhance power systems’ extreme weather resilience,'' \emph{Journal of Infrastructure Systems}, vol.~27, no.~2, 2021.

\bibitem{sturmer2024increasing}
J.~St{\"u}rmer, A.~Plietzsch, T.~Vogt, F.~Hellmann, J.~Kurths, C.~Otto, K.~Frieler, and M.~Anvari, ``Increasing the resilience of the {T}exas power grid against extreme storms by hardening critical lines,'' \emph{Nature Energy}, vol.~9, no.~5, pp. 526--535, May 2024.

\bibitem{bilir2024enhancing}
B.~Bilir, E.~Kutanoglu, J.~J. Hasenbein, B.~Austgen, M.~Garcia, and J.~K. Skolfield, ``Enhancing power grid resilience to winter storms via generator winterization with equity considerations,'' \emph{Sustainable Cities and Society}, vol. 114, 2024.

\bibitem{austgen2024three}
B.~G. Austgen, M.~Garcia, J.~J. Yip, B.~Arguello, B.~J. Pierre, E.~Kutanoglu, J.~J. Hasenbein, and S.~Santoso, ``Three-stage optimization model to inform risk-averse investment in power system resilience to winter storms,'' \emph{IEEE Access}, vol.~12, pp. 135\,117--135\,134, 2024.

\bibitem{pollack2024equitably}
M.~Pollack, R.~Piansky, S.~Gupta, and D.~K. Molzahn, ``Equitably allocating wildfire resilience investments for power grids -- {T}he curse of aggregation and vulnerability indices,'' \emph{Applied Energy}, vol. 388, no. 125511, June 2025.

\bibitem{piansky2024long}
R.~Piansky, G.~Stinchfield, A.~Kody, D.~K. Molzahn, and J.~P. Watson, ``Long duration battery sizing, siting, and operation under wildfire risk using progressive hedging,'' \emph{Electric Power Systems Research}, vol. 235, no. 110785, October 2024, presented at the \textit{23rd Power Systems Computation Conference (PSCC)}.

\bibitem{skolfield2021transmission}
J.~K. Skolfield, J.~Ramirez-Vergara, and A.~R. Escobedo, ``Transmission and capacity expansion planning against rising temperatures: A case study in {A}rizona,'' \emph{arXiv:2106.12687}, 2021.

\bibitem{nasri2022multi}
A.~Nasri, A.~Abdollahi, and M.~Rashidinejad, ``Multi-stage and resilience-based distribution network expansion planning against hurricanes based on vulnerability and resiliency metrics,'' \emph{International Journal of Electrical Power \& Energy Systems}, vol. 136, p. 107640, 2022.

\bibitem{newlun2019co}
C.~J. Newlun, A.~L. Figueroa-Acevedo, J.~D. McCalley, A.~Kimber, and E.~O'Neill-Carrillo, ``Co-optimized expansion planning to enhance electrical system resilience in {P}uerto {R}ico,'' in \emph{51st North American Power Symposium (NAPS)}, 2019.

\bibitem{moreira2021climate}
A.~Moreira, D.~Pozo, A.~Street, E.~Sauma, and G.~Strbac, ``Climate-aware generation and transmission expansion planning: A three-stage robust optimization approach,'' \emph{European Journal of Operational Research}, vol. 295, no.~3, pp. 1099--1118, 2021.

\bibitem{CPUCignitions}
\BIBentryALTinterwordspacing
{California Public Utilities Commission}, ``Fire ignition data,'' 2023. [Online]. Available: \url{www.cpuc.ca.gov/industries-and-topics/wildfires}
\BIBentrySTDinterwordspacing

\bibitem{sotolongo2020california}
M.~Sotolongo, C.~Bolon, and S.~H. Baker, ``California power shutoffs: Deficiencies in data and reporting,'' \emph{Initiative for Energy Justice}, October 2020, \url{https://iejusa.org/wp-content/uploads/2020/10/V3.3-Policy-Brief-CA-Shutoffs-Data-Brief.pdf}.

\bibitem{bayani2023resilient}
R.~Bayani and S.~D. Manshadi, ``Resilient expansion planning of electricity grid under prolonged wildfire risk,'' \emph{IEEE Transactions on Smart Grid}, vol.~14, no.~5, 2023.

\bibitem{Taylor2023Managing}
S.~Taylor, G.~Setyawan, B.~Cui, A.~Zamzam, and L.~A. Roald, ``Managing wildfire risk and promoting equity through optimal configuration of networked microgrids,'' in \emph{ACM e-Energy}, June 2023.

\bibitem{kody2022optimizing}
A.~Kody, R.~Piansky, Molzahn, and D.~K, ``Optimizing transmission infrastructure investments to support line de-energization for mitigating wildfire ignition risk,'' \emph{11th IREP Symposium on Bulk Power System Dynamics and Control}, 2022.

\bibitem{taylor2023robust}
S.~Taylor, A.~Musselman, L.~Roald, and J.~Watson, ``A robust optimization approach to determine power line undergrounding under wildfire risk,'' Lawrence Livermore National Laboratory (LLNL), Livermore, CA (United States), Tech. Rep., 2023.

\bibitem{taylor2022framework}
S.~Taylor and L.~A. Roald, ``A framework for risk assessment and optimal line upgrade selection to mitigate wildfire risk,'' \emph{Electric Power Systems Research}, vol. 213, 2022, presented at the \textit{22nd Power Systems Computation Conference (PSCC)}.

\bibitem{xu2017creation}
T.~Xu, A.~B. Birchfield, K.~S. Shetye, and T.~J. Overbye, ``Creation of synthetic electric grid models for transient stability studies,'' in \emph{10th Bulk Power Systems Dynamics and Control Symposium (IREP)}, 2017.

\bibitem{brelsford2024dataset}
C.~Brelsford, S.~Tennille, A.~Myers, S.~Chinthavali, V.~Tansakul, M.~Denman, M.~Coletti, J.~Grant, S.~Lee, K.~Allen \emph{et~al.}, ``A dataset of recorded electricity outages by {U}nited {S}tates county 2014--2022,'' \emph{Scientific Data}, vol.~11, no.~1, p. 271, 2024.

\bibitem{noaa2024sed}
\BIBentryALTinterwordspacing
{National Climatic Data Center, NESDIS, NOAA, U.S. Department of Commerce}, ``Storm events database, version 3.1,'' 2024. [Online]. Available: \url{https://www.ncdc.noaa.gov/stormevents/}
\BIBentrySTDinterwordspacing

\bibitem{huang2023review}
C.~Huang, Q.~Hu, L.~Sang, D.~D. Lucas, R.~Wong, B.~Wang, W.~Hong, M.~Yao, and V.~Donde, ``A review of public safety power shutoffs {(PSPS)} for wildfire mitigation: Policies, practices, models and data sources,'' \emph{IEEE Transactions on Energy Markets, Policy and Regulation}, vol.~1, no.~3, pp. 187--197, 2023.

\bibitem{xu2018synthetic}
T.~Xu, A.~B. Birchfield, and T.~J. Overbye, ``Modeling, tuning, and validating system dynamics in synthetic electric grids,'' \emph{IEEE Transactions on Power Systems}, vol.~33, no.~6, pp. 6501--6509, 2018.

\bibitem{mahmood3analysis}
R.~S. Mahmood, R.~J. Mizban, M.~A. Sarhan, A.~Rashid, M.~Rasheed, and T.~Saidani, ``Analysis and applications of the beta prime distribution in statistical modeling,'' \emph{Journal of Positive Sciences}, vol.~3, no.~6, pp. 34--41.

\bibitem{mahmood3utilizing}
------, ``Utilizing beta distribution for probabilistic modeling: Five numerical examples,'' \emph{Journal of Positive Sciences}, vol.~3, no.~5, pp. 40--48.

\bibitem{rts96}
C.~Grigg, P.~Wong \emph{et~al.}, ``{The IEEE Reliability Test System-1996. A Report Prepared by the Reliability Test System Task Force of the Application of Probability Methods Subcommittee},'' \emph{IEEE Transactions on Power Systems}, vol.~14, no.~3, pp. 1010--1020, 1999.

\bibitem{USGS2024Wildland}
{U.S. Geological Survey}, ``{Wildland Fire Potential Index},'' 2024.

\end{thebibliography}

\end{document}